# PSO-RBF Based control Schema for Adaptive Active Queue Management in TCP Networks

Mansour Sheikhan[1], Reza Shahnazi[2], Ehasn Hemmati[3]

1- Department of Communication Engineering, Islamic Azad University, South Tehran Branch, Tehran, Iran{msheikhn@azad.ac.ir}

2- Modeling and Optimization Research Center in Science and Engineering, Islamic Azad University, South Tehran Branch, Tehran, Iran {shahnazi@ieee.org}

3- Department of Electronic Engineering, Islamic Azad University, South Tehran Branch, Tehran, Iran{ehemmati@ieee.org }

Address: Faculty of Engineering, Islamic Azad University, South Tehran Branch, Tehran, Iran,

**Abstract**

Addressing performance degradations in end-to-end congestion control has been one of the most active research areas in the last decade. Active queue management (AQM) aims to improve the overall network throughput, while providing lower delay and reduce packet loss and improving network. The basic idea is to actively trigger packet dropping (or marking provided by explicit congestion notification (ECN)) before buffer overflow. Radial bias function (RBF)-based AQM controller is proposed in this paper. RBF controller is suitable as an AQM scheme to control congestion in TCP communication networks since it is nonlinear. Particle swarm optimization (PSO) algorithm is also employed to derive RBF parameters such that the integrated-absolute error (IAE) is minimized. Furthermore, in order to improve the robustness of RBF controller, an error-integral term is added to RBF equation. The results of the comparison with Drop Tail, adaptive random early detection (ARED), random exponential marking (REM), and proportional-integral (PI) controllers are presented. Integral-RBF has better performance not only in comparison with RBF but also with ARED, REM and PI controllers in the case of link utilization while packet loss rate is small.

# 1. Introduction

In recent years, the unpredictable growth of the Internet has increasingly pointed out the network traffic congestion problem. Congestion at routers may be caused by too many sources trying to send an excessive amount of data with a rate that is too high for the network to handle. Network traffic congestion results in long time delays for data transmission and frequently makes the queue length in the buffer of the intermediate router overflow, and can even lead to network collapse [1, 2]. Buffer management for Internet routers plays an important role in congestion control. Dropping the packets serves as a critical mechanism of congestion notification to the end nodes.

Transmission control protocol (TCP) is a connection-oriented transport layer protocol of the Internet. One of the TCP features is the cutting back on the transmission rate of flows whenever a congestion is occurred along the path of the packet flow. A packet loss could mean that one of the intermediate routers does not have enough buffer space to store the packet before its transmission on the appropriate link toward the destination. A small buffer generally achieves a low queuing delay, but suffers from excessive packet losses and low link utilization, and vice versa.

Active queue management (AQM), which is a proactive approach, has been proposed as a solution to these problems [4]. AQM policies are those policies of router queue management that allow for the detection of network congestion, the notification of such occurrences to the hosts, and the adoption of a suitable control policy. The idea behind AQM is the early notification of incipient congestion so that TCP senders can reduce their transmission rate before queue overflows. A simple policy like the widely used First-In-First-Out (FIFO) Tail-Drop often causes strong correlations among packet losses, resulting in the well-known "TCP synchronization" problem [101]. AQM policies based on control theory consider the intrinsic feedback nature of congestion systems. Sources adjust their transmission rates according to the level of congestion. The notification of congestion to the sources is done through packet dropping or marking, based on AQM policy. The term "marking" uses more generally to refer to any action taken by the AQM algorithm to notify the source of incipient congestion. Controllers determine the appropriate probability of the packet dropping or marking. An AQM mechanism can reduce the number of dropping packets as well as the delays seen by flows by keeping small the average queue size [5, 6]. In recent years, several different approaches have resulted in various AQM policies [7-10]. AQM policies, when properly used, provide better network utilization and lower end-to-end delays than drop tail. The development of new AQM routers will play a key role in meeting tomorrow's increasing demand for performance in Internet applications. Such applications include voice over IP (VoIP), class of service (CoS), and streaming video where the packet size and session duration exhibit significant variations.

Random early detection (RED) algorithm is the earliest and the most prominent of AQM schemes [4] which is recommended by the Internet Engineering Task Force (IETF), to be deployed in the Internet. The notify the traffic sources by early dropping or marking of the packets, to be able to avoid the global synchronization problem, maintaining low average queuing delay and providing better utilization. However, the behavior of RED strongly depends on tuning parameters for every specific case and average queue size varies with the level of congestion. As a result, the average queuing delay from RED is sensitive to the traffic load and also to parameters setting, and is therefore not predictable in advance. Delay is a major component of the quality of service delivered to the customers, so network operators would naturally like to have a rough a priori estimate of the average delays in their congested routers. To achieve such predictable average delays with RED, we would require constant tuning of RED's parameters to adapt with the current traffic conditions. A second, related weakness of RED is that the throughput is also sensitive to the traffic load and to the RED parameters. Avoiding this regime would again require constant tuning of the RED parameters.

Due to these drawbacks some new modified schemes including adaptive RED (ARED) [11], Flow Random Early Detection (FRED) [12], stabilized RED (SRED) [13], balanced RED (BRED) [14], BRED with virtual buffer occupancy (BRED/VBO) [15], exponential RED (ERED) [16], dynamic RED (DRED)

[17], novel autonomous proportional and differential RED (PD-RED) [18, 102], loss ratio and rate control RED (LRC-RED) [19], robust RED (RRED) [20], Neural network based RED (NN-RED) [103], random exponentially marking (REM) [6], adaptive virtual queue (AVQ) [5], BLUE [21], YELLOW [22], and GREEN [23] have been proposed as alternatives to RED.

In this paper, neural-based control algorithms are proposed for AQM with the optimized parameters using particle swarm optimization (PSO) algorithm. The neural models are radial basis function (RBF) and a modified RBF which increases the robustness of the controller. It is noted that RBF is an artificial neural network whose hidden layers are composed by radial basis function. RBFs were first introduced in the solution of the real multivariate interpolation problem, and it is now being extensively investigated as a branch of neural networks for interpolation and classification [24]. RBF is neither a local nor a global interpolation function. Its spatial support range is moderate and tunable [25]. As mentioned above, we propose PSO-optimized RBF-based controllers for queue management in this paper. Due to nonlinear characteristic of the traffic in the network, RBF is suitable to control the queue and achieve desired quality of service (QoS). As a modified RBF model, we also add an error-integral term to RBF equations to increase the robustness and improve the performance of AQM controller. This modified RBF model is called IRBF in this paper.

PSO, as a heuristic optimization algorithm which is developed through simulation of a simplified social system, is used for solving the optimal proposed RBF controllers' parameters problem. The PSO technique can generate a high-quality solution with short calculation time and more stable convergence characteristic than other stochastic methods [26].

Two main contributions are considered in this paper. First, general nonlinear TCP model including delay is presented. Second, we propose PSO-RBF and PSO-IRBF AQM controllers based on this TCP model.

The rest of the paper is organized as follow. Related work is reviewed in Section 2. In section 3, the TCP model is described, and RBF neural model and foundation of PSO algorithm are introduced. The proposed methods based on PSO-optimized RBF and IRBF controllers are illustrated in Section 4. Section 5 shows the simulation results and parameter settings. Finally, the work is concluded in Section 6.

## 2. Related Work

An AQM algorithm controls the network congestion by randomly dropping or marking packets at the router queue. The process of reducing the transmission rate by the TCP sources and decreasing the queue size constitutes a closed loop feedback control system where the controlled variable is the queue size and the control variable is the dropping/marking probability. In this way, different control strategies have been used for AQM such as proportional-integral (PI)-based control [27-30], proportional-differential (PD)- based control [31], proportional-integral-differential (PID)-based control [32-35], adaptive control [36], fuzzy control [37-39,106, 108], optimal control [40], predictive functional control (PFC) [41], static state feedback control [42], observer-based control [42], variable structure control (VSC) [43], sliding mode control [44], robust control [45, 46], virtual rate control (VRC) [47], $H_\infty$ approach [10, 48, 49], and neural-based control [8, 50, 51, 103].

As examples of proposed PI and modified PI AQM controllers, Hollot et al. [27] have developed a PI controller to regulate the queue level, round trip time (RTT), and packet loss. A Smith predictor-based PI (SPPI) controller has been proposed by Li et al. [28] to reduce the effects of time delays in the control loop. Chang and Muppala [29] have proposed a stable queue-based adaptive PI (S-QAPI) controller for AQM to improve the transient performance of the fixed-gain PI controller over a wide range of uncertainties. It is noted that in the PI controller, the usage of a proportional section leads to a lower time of response but also to lower stability margins. Moreover, the proportional controller has a steady state regulation error. In order to overcome these disadvantages, the integral term is added which has the characteristic to give steady state error equal to zero and to give higher stability margins [13, 34]. However, the PI-based AQM suffers the disadvantage that it cannot maintain good dynamic performance as the number of TCP flow increases.

As an example of PD-based AQM controller, Kim and Low [31] have formulated the problem for stabilizing a given TCP network described by state-space models and PD controller.

As examples of proposed PID and modified PID AQM controllers, Ren et al. [32] have presented an AQM algorithm based on PID control for large delay networks. On the other hand, several artificial intelligence (AI)-based techniques have been proposed to improve the controller performances, such as genetic algorithm (GA) [34], PSO [33], and neural network [35].

As an example of adaptive control scheme for AQM, a self-tuning algorithm called adaptive queue control (AQC) has been proposed in [36] which exploits online estimates of the network parameters.

As examples of fuzzy control schemes for AQM, Di Fatta et al. [37], and Fengyua and Xiuming [38] have proposed congestion control algorithms based upon fuzzy logic controllers. Also, a fuzzy adaptive AQM controller (FAAQMC) has been proposed in [39] which suppresses buffer overflow by adapting the buffer size to the required queue length with a control cycle time shorter than the mean inter-arrival time of a burst. LIU et al. proposed a fuzzy congestion control algorithm based on fuzzy logic which uses the pre dominance of fuzzy logic to deal with uncertain events [106]. The main advantage of this algorithm is that it discards the packet dropping mechanism of RED, and calculates packet loss according to a preconfigured fuzzy logic by using the queue length and the buffer usage ratio, which enables to obtain improved performance under dynamic environments. In this algorithm queue length can be kept stable in a variety of network environment without difficulty of parameter configuration. An adaptive fuzzy sliding mode (AFSM) AQM [108] is presented by X. Guan et al. The AFSM algorithm uses the queue length and its differential as the input of AQM and adjusts fuzzy rules by the measurement of packet loss ratio dynamically. Because of stability analysis under heterogeneous round-trip time (RTT) and re-adjustment of the fuzzy rules, the stability of AFSM is independent of the active flow numbers.

As an example of optimal control scheme for AQM, a on-rational approach has been proposed in [40] in which stability and performance objectives of the system have been completely expressed as linear matrix inequalities (LMIs).

As an example of predictive control scheme, Bigdeli and Haeri [41] have proposed a simple and low-computational load PFC-AQM method for high speed networks.

As other examples of classic control approaches for AQM, Chen et al. [42] have proposed static state feedback control and observer-based control laws and delay-independent stability criteria have been derived by applying the Lyapunov-Krasovskii functional approach and the LMI technique. The VSC-based AQM controller has been also presented in [43]. Fengyuan et al. [44] have proposed the sliding mode variable structure control which is based on robust control theory to achieve robustness against disturbances and variance of parameters. As another robust controller for AQM, a congestion control law has been obtained in [45] through an interactive loop-shaping process that manipulates the system frequency response to meet robust stability and performance requirements in the presence of uncertain network conditions. Similarly, Bigdeli and Haeri [46] have proposed an AQM controller based on coefficient diagram method (CDM) which is an indirect pole placement method that considers the speed, stability and robustness of the closed loop system in terms of time-domain specifications. Besides, an adaptive CDM (ACDM) has been developed in which the output feedback pole placement is implemented in an indirect characteristic polynomial is determined by CDM and the system parameters are estimated using a modified recursive gradient method.

As examples of the cotroller for AQM, Quet and Ozbay [48] have applied $H_\infty$ aproach to AQM using the linearized TCP model. Yu et al. [10] have proposed a robust controller for AQM based on modern $H_\infty$ optimal control theory with parameter tuning. In addition, Manfredi et al. [49] have developed a robust $H_\infty$ controller for time-delay systems to cope with unwanted variations of characteristic parameters such as average RTT, the load and link capacity.

As examples of neural-based controllers for AQM, Rahnami et al. [50] have implemented a neural network-based model reference control (MRC) algorithm to improve transient and steady state behavior of TCP flows and AQM routers. In this way, two neural networks have been trained to control the traffic flow of a bottleneck router. The mentioned neural networks have been used to model and control the system, respectively. The model network has been trained offline. Then, the controller has been trained such that system response follows a reference model. Each network had two layers with delayed inputs and outputs. Two and three delayed inputs and outputs were selected, respectively. Cho et al. [8] have proposed a multi-layer perceptron-infinite impulse response (MLP-IIR) as a recurrent neural network (RNN) AQM controller. In this way, three distinct neural networks have been trained under different network scenarios involving traffic levels. Selecting each of three mentioned neural AQMs is based on posterior probability history of traffic level. NN-RED algorithm is a random early detection scheme that takes advantage of a neural network predict future values of queue size based on current and previous values of the queue length [103]. Recently, Lochin and Talavera [51] have configured the RED parameters by using a Kohonen neural network model which enables a stable queue length with complex parameters setting.

Zhenyu et al. proposed an adaptive traffic prediction based AQM (AT-PAQM) algorithm [104] . NOEKF, which is the combination of the Kalman filtering model and the online noise estimation, accurately predict traffic. NOEKF is used to predict congestion indication in ATPAQM algorithm. The packet loss ratio and the packet dropping probability are calculated in both coarse and fine granularities based on the prediction results. In coarse granularity, traffic is predicted every prediction interval, then the average packet loss ratio is calculated. In fine granularity, upon a packet arriving, the packet dropping probability is adjusted according to the average packet loss ratio.

Zhang et al. introduced an optimal controller to improve AQM router's stability and response time based on Lyapunov function [105]. Their algorithm has better performance in response time and stability than PI controller but it is designed only for a single link.

An AQM controller for multilayer network is presented in [107]. It is based on theory which considers that when there are many edge routers, the stability of the fluid model ensures the convergence of the average rate to its equilibrium point, and may impose excessive restrictions on the choice of system parameters. It is based on heuristics and data traffic controllers which adjust the network parameters with the multilayer network status.

In this paper, we propose two RBF-based AQM controllers which their parameters are obtained optimally using PSO algorithm.

## 3. Problem Formulation and Preliminaries

*3.1. TCP/AQM model based on fluid flow theory*

A dynamic model of TCP behavior has been developed using fluid flow and stochastic differential equation analysis in [52]. Briefly, this model expresses the coupled nonlinear differential equations such that they reflect the dynamics of TCP accurately with the average TCP window size and the average queue length. A simplified version of this model, which does not include the TCP timeout mechanism, is given by [30]:

$$\dot{w}(t) = \frac{1}{\frac{q(t)}{C}+T_p} - \frac{w(t)}{2}\frac{w(t-R(t))}{\frac{q(t-R(t))}{C}+T_p} p(t-R(t)) \quad (1)$$

$$\dot{q}(t) = \begin{cases} -C + \dfrac{N(t)}{\frac{q(t)}{C}+T_p} w(t) & \text{if } q(t)>0 \\[1em] \max\left\{0, -C + \dfrac{N(t)}{\frac{q(t)}{C}+T_p} w(t)\right\} & \text{if } q(t)=0 \end{cases} \quad (2)$$

where $w(t)$ is the mean TCP window size (in packets), $q(t)$ is the queue size (in packets), $R(t)$ is the RTT (in seconds) and equals to $q/C+T_p$, $C$ is the link capacity (in packets/second), $N(t)$ is the number of TCP connections, and $p(t)$ is the packet mark/drop probability.

Equation (1) describes the TCP window dynamics. The first term models the window additive increase, while the second term models the multiplicative decrease. Equation (2) expresses the queue size variation as the difference between the arrival rate, $NW/R$, and the link capacity, $C$.

Since the packet-dropping probability is between 0 and 1, the following nonlinear time-delayed system with a saturated input can be derived:

$$\dot{w}(t) = \frac{1}{\frac{q(t)}{C}+T_p} - \frac{w(t)}{2}\frac{w(t-R(t))}{\frac{q(t-R(t))}{C}+T_p} sat(u(t)) \quad (3)$$

$$\dot{q}(t) = \begin{cases} -C + \dfrac{N(t)}{\frac{q(t)}{C}+T_p} w(t) & \text{if } q(t)>0 \\[1em] \max\left\{0, -C + \dfrac{N(t)}{\frac{q(t)}{C}+T_p} w(t)\right\} & \text{if } q(t)=0 \end{cases} \quad (4)$$

The saturated input $u(t) = p(t-R(t))$ is expressed by the following nonlinearity:

$$sat(u(t)) = \begin{cases} u_{max}, & u(t) \geq u_{max} \\ u(t), & u_{min} \leq u(t) < u_{max} \\ u_{min}, & u(t) < u_{min} \end{cases} \quad (5)$$

where the lower boundary and upper boundary are given as $u_{min}=0$ and $u_{max}=1$.

## 3.2. RBF neural network

The RBF neural networks result from the use of radial basis functions in the solution of the real multivariate interpolation problem [53, 54]. RBF network can be used in a wide range of applications primarily because it can approximate any regular function. As shown in Fig. 1, basic form of a RBF neural network involves three different layers. The input layer is the set of source nodes connected to the second layer, which is a hidden layer of high dimension. The output layer gives the response of the network. The transformation from the input space to the hidden-unit space is nonlinear. On the other hand, the transformation from the hidden space to the output space is linear.

The output, $y$, of a Gaussian RBF network is evaluated from the input vector, $x$, as follows:

$$y(x) = w^T \exp(\frac{-\|x-c\|^2}{r^2}) \tag{6}$$

where $y(x)$ is approximate function, $w$ is weight for basis function. $c$ and $r$ are the center and spread of Gaussian function.

## 3.3. PSO algorithm

PSO is a population based stochastic optimization technique introduced by Eberhart and Kennedy [55] inspired by social behavior of bird flocking or fish schooling, specially the ability of birds to flock in to search of food. This behavior is associated to that of an optimization search for the solution to nonlinear equation in a search space. PSO does not use the gradient of problem being optimized, so it does not require being differentiable for the optimization problem as is required in classic optimization algorithms. Therefore it can also be used in optimization problems that are partially irregular, time variable, and noisy.

Each bird, referred to as a 'particle', represents a possible solution for the problem. Each particle moves through the D-dimensional problem space by updating its velocities with the best solution found by itself (cognitive behavior) and the best solution found by any particle in its neighborhood (social behavior). In PSO, particles move in a multidimensional search space. In this algorithm, each particle has a velocity and a position as follow:

$$v_i(k+1) = v_i(k) + \gamma_{1i}(P_i - x_i(k)) + \gamma_{2i}(G - x_i(k)) \tag{7}$$

$$x_i(k+1) = x_i(k) + v_i(k+1) \tag{8}$$

where $i$ is the particle index, $k$ is the discrete time index, $v_i$ is the velocity of $ith$ particle, $x_i$ is position of $ith$ particle, $P_i$ is the best position found by ith particle (personal best), $G$ is the best position found by swarm (global best) and $\gamma_{1,2}$ are random numbers in the interval [0,1] applied to *ith* particle. In our simulations, the following equation is used for velocity [56]:

$$v_i(k+1) = \varphi(k)v_i(k) + \alpha_1\left[\gamma_{1i}(P_i - x_i(k))\right] + \alpha_2\left[\gamma_{2i}(G_i - x_i(k))\right] \quad (9)$$

in which $\phi$ is inertia function and $\alpha_{1,2}$ are the acceleration constants.

## 4. Proposed PSO-RBF Methods for AQM

*4.1. RBF-based AQM scheme*

In this section, the RBF controller is proposed to achieve the desired queue length efficiently with delay effects and a saturated input. An RBF controller generates the term $u(t)$ as a control input in (3). The output error signal is defined as $e(t) = q(t) - q_t$, where $q_t$ denotes the target queue length. In this application, a Gaussian RBF controller with an input $e(t)$ and an output $u(t)$ is expressed as follows:

$$u(t) = u = w^T \varphi(e) \quad (10)$$

where $w$ is the weight of hidden layer of RBF network and $\varphi$ is defined as follows:

$$\varphi_i(e) = \exp\left(-\frac{(\|e - c_i\|)^2}{\sigma_i^2}\right) \quad (11)$$

where $c_i$ is the mean and $\sigma_i$ is the spread of *i*th radial bias function.

In order to measure the performance of the closed-loop control system, an integral absolute error (IAE) measure is employed with the following equation:

$$IAE = \frac{1}{T}\int_0^T |e(\tau)| d\tau \quad (12)$$

In fact, the integral absolute error will depend on the controller parameters which here are the weight of hidden layer, means and spreads. If the smaller value of IAE is achieved, then the better RBF controller is designed. Therefore the goal is to find the optimum values of RBF parameters which made the IAE minimum. Practically, the design of the RBF's parameters is very heuristic and depends on the expert's experiences. The proposed RBF-based controller schema is depicted in Fig. 1. In the next section, a PSO-based method determining the optimal weight of RBF network to minimize the IAE will be presented.

*4.2. IRBF-based AQM scheme*

In order to improve the controller performance, we add an error-integral term to the proposed RBF-based controller. So, the relation between controller input *e(t)* and the control signal *u(t)* is defined as follows:

$$u(t) = u = w^T \varphi(e) + w_I \int_0^t e(\tau)\, d\tau \qquad (13)$$

where $w_0$ is the integral gain and $w$ is the weight of hidden layer of RBF network and $\varphi$ is defined as (11).

Again IAE is employed in order to measure the performance of the closed-loop control system. In fact, the integral absolute error will depend on the controller parameters which here are the weight of hidden layer, means and spreads. Similarly, if the smaller value of IAE is achieved, then the better RBF controller is designed. Hence, the goal is to find the optimum values of RBF parameters and $w_I$ such that the IAE becomes minimum. PSO algorithm is used in order to tune the controller parameters since it has good performance in continuous optimization problems. By using PSO to tune the parameters there is no need to define the parameters manually which is very heuristic and depends on expert's experiences. The controller parameters are defined according to the target function, IAE, which makes the parameters optimum since PSO tries to minimize this target function. The proposed IRBF-based controller scheme is depicted in Fig. 2.

## 5. Simulation Results

*5.1. Parameter tuning in controller design*

In this section, parameter setting of the RBF- and Integral-RBF in controller design for TCP/AQM is illustrated. For a TCP/AQM network modeled by (3) and (4), it is assumed that N = 100 homogeneous TCP connections share one bottleneck link with a capacity of 10 Mbps; i.e., $C$ = 1250 (packets/second). Furthermore, the propagation delay of the bottleneck link capacity is set to $T_p$ = 60 msec, and the desired queue size is assumed as $q_d$ = 150 packets. The time duration of queue monitoring, $T$, is set to 100 seconds. The initial values of PSO parameters are set according to values depicted in Table 1. The IAE value is obtained as an average over 11 simulations using 11 initial queue size uniformly distributed between 0 and 300 in each of iterations. One of the most important settings in this design is the number of hidden layer neurons. We use PSO algorithm for several RBF networks with different number of neurons in hidden layer. The parameter values and IAE value results from PSO is shown in Table 2. Variance of all functions is set to 40 and the mean is distributed between -150 and 150 steps by 75.

Table 1. Initial values of PSO Parameters

| Parameter | Maximum number of iterations | Population size | Maximum particle velocity | Initial inertia weight | Final inertia weight | Minimum global error gradient |
|---|---|---|---|---|---|---|
| Value | 300 | 20 | 4 | 0.9 | 0.2 | $10^{-5}$ |

Table 2. IAE values for five different RBF-based controllers

| Number of radial basis functions | 3 | 5 | 7 | 9 | 11 |
|---|---|---|---|---|---|
| IAE | 0.462 | 0.381 | 0.385 | 0.374 | 0.373 |

Since the number of parameters is increased as the number of radial basis functions is increased, the proposed neural controllers consist of 5 radial basis functions for both RBF controller and Integral-RBF controller. Table 3 shows the PSO output for optimal parameter setting of RBF and Integral-RBF controller. The convergence curve of the IAE value against number of iterations for these controllers is shown in Fig. 3.

Table 3. PSO-optimized parameters and IAE of proposed AQM controllers

| AQM scheme | Optimal parameters | IAE |
|---|---|---|
| RBF | $w_1=-1$, $w_2=-1$, $w_3=0.3397$, $w_4=0.3372$, $w_5=1$ | 0.3627 |
| Integral-RBF | $w_1=-1$, $w_2=-0.9612$, $w_3=0.3445$, $w_4=0.9939$, $w_5=0.9979$, $w_I=7.0813\times10^{-4}$ | 0.3832 |

*4.2. Performance evaluation for constant TCP connections*

The performance and effectiveness of the proposed RBF-based controllers have been verified in a series of numerical simulations using NS-2 (Network Simulator-2) with the dumbbell network topology, which is shown in Fig. 4. In dumbbell network topology, the transport agent is based on TCP-Reno, where multiple TCP connections share a single bottleneck link. Each link capacity and the propagation delay, is also depicted in Fig. 4. It is supposed that the TCP sources send their data incessantly. Unless otherwise noted, the number of connections is assumed to be constant and equal to 100 in our simulation. The maximum buffer size of each router is assumed to be 300 packets; each packet has a size of 1000 bytes. The desired queue length, *qd*, is set to 150 packets. To demonstrate the robustness of the proposed AQM controller, the dynamic traffic changes and different RTTs in the simulated network are taken into account in the simulations. The simulation results for these conditions using the proposed control strategy are compared with other AQM schemes, such as Drop tail, PI [27], REM [6], and ARED [57]. The parameters of mentioned AQM schemes are listed in Table 4.

Table 4. Parameter setting of PI, REM and ARED controllers

| AQM scheme | Parameter setting |
|---|---|
| PI | $a = 1.822 \times 10^{-5}$, $b = 1.816 \times 10^{-5}$, $T = 1/160$ s |
| REM | $\gamma = 0.001$, $\varphi = 1.001$ |
| ARED | $min_{th} = 100$, $max_{th} = 215$, $w_g = 1-\exp(-1/C)$ |

Figure 5 shows the simulation results for mentioned controllers. RBF and Integral-RBF are converged to the desired queue size more rapidly than other mentioned AQM schemes. Although responses of PI and ARED can achieve and maintain the queue length around the desired value, too inactive and serious overshoots occurred. The REM scheme is the worst case, since there is no parameter of the target queue length in its control algorithm, so it cannot maintain the desired queue length.

*5.3. Performance evaluation for dynamic traffic load*

In this scenario, performance of different AQM schemes, considering dynamic network traffic, are evaluated. At the beginning, t=0, 100 TCP-Reno connections are established. At t=30, 30 more TCP-Reno connections begin transmission and remained inactive until time t = 60. Additionally, 30 TCP connections departed at the same time so there is only 70 active connections till t = 80, which 100 connections would transmitt till the end of the simulation period. Fig. 6 shows the corresponding queue evolutions obtained under the various AQM schemes. It can be seen that Drop Tail, PI , ARED and REM controllers are not robust with respect to variations in the load. The proposed RBF and Integral-RBF controllers are robust to variations in the number of active connections.

*5.4. Performance evaluation in short and long propagation delays*

The robustness of the proposed methods against variations of the RTT is evaluated in this scenario. At first, assume that the links between senders and $R_1$ and also between $R_2$ and receivers are characterized as bandwidth of 10 Mbps and a short inherent propagation delay of 2 msec. There is a bandwidth of 10 Mbps and an inherent propagation delay of 10 msec between $R_1$ and $R_2$. The responses of the queue length obtained with different AQM schemes are shown in Fig. 7. The effect of a long inherent propagation delay is also considered, where the TCP sources and destinations are linked respectively to $R_1$ and $R_2$ with a bandwidth of 10 Mbps and an inherent propagation delay of 20 msec. An inherent propagation delay of 140 msec between $R_1$ and $R_2$ is chosen in this case. The responses of the queue length obtained with different schemes are shown in Fig. 8. ARED has a steady-state error with a short inherent propagation delay and has serious oscillation when given a long inherent propagation delay. Again Drop tail is not robust and shows periodic behavior. Even though PI has no steady-state error in the network with long propagation delay but, the transient responses are too sluggish. RBF has small steady-state error, but Integral-RBF achieved the desired queue length in a reasonable transient response time.

*5.5. Performance evaluation of link utilization and packet loss rate*

The responses of the link utilization and packet loss rate in the different number of users and different bottleneck link propagation delay are also studied through simulations. In this case, the number of users is varied between 70 and 160 and different propagation delays with diverse times of between 20 and 140 msec. In Figs. 9 and 10, the link utilization and packet loss rate in different number of users are depicted, respectively. Link utilization of the Integral-RBF model is higher than other AQM controllers except for 160 TCP connections and it is very close to the ARED scheme. Furthermore, for all AQM controllers, the link utilization is raised when the number of users is increased. As shown in Fig. 10, the Integral-RBF controller has loss rate a little more compared with the PI and ARED schemes. But the difference between RBF controller and the others is almost noticeable. Link utilization of Drop Tail is almost constant and it is less than all AQM schemas except REM, on the other hand the packet loss of Drop tail is also less as it uses whole queue length. Figures 11 and 12 depict the responses of the link utilization and packet loss rate in the different propagation delays, respectively. Drop tail link utilization is almost less than AQM schemas and its packet loss is also less but it increases as delay is icreased. The performance of PI, ARED, RBF and Integral RBF is very close to each other and in all of them link utilization is decreased as the delay increases. Packet loss for Integral-RBF is higher than PI and ARED but it is very close to them, since the link utilization is better so it can be neglected in this case.

# 6. Conclusion

Robust active queue management schemes at the routers are essential for communication network. In this paper, we have proposed a RBF-based AQM controller to enhance TCP congestion control. In order to improve the robustness of proposed method we have also introduced an Integral-RBF AQM controller. PSO algorithm has been used in order to tune the parameters of mentioned controllers in such a way that the integrated-absolute error (IAE) is minimized. A deep analysis of the RBF and I-RBF stability and performance has been carried out through simulations using NS-2 tool. The experimental results have demonstrated that the I-RBF controller outperforms the other AQM policies under various operating conditions. It achieves both good queue regulation and high link utilization. It is also shown that the proposed controller has fast response and also is robust to high load variations and disturbance rejection in steady-state behavior. Link utilization of the Integral-RBF is higher than Drop Tail, PI, REM and ARED controllers while packet loss is small and very close to mentioned controllers.

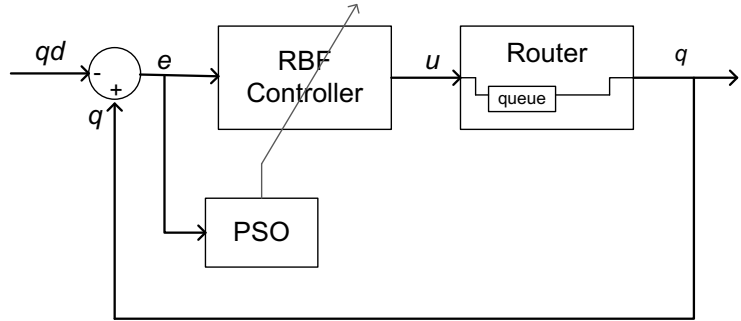

Fig. 1. PSO-optimized RBF-based controller for AQM

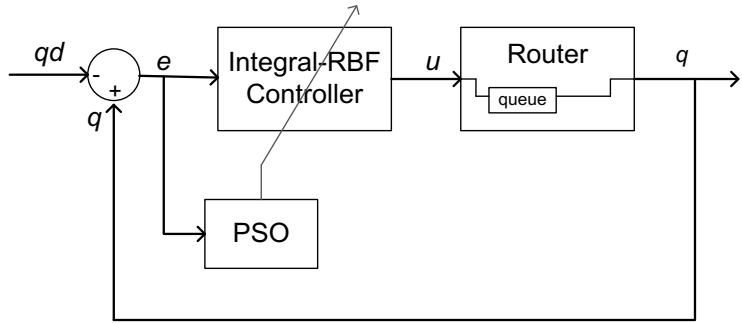

Fig. 2. PSO-optimized IRBF-based controller for AQM

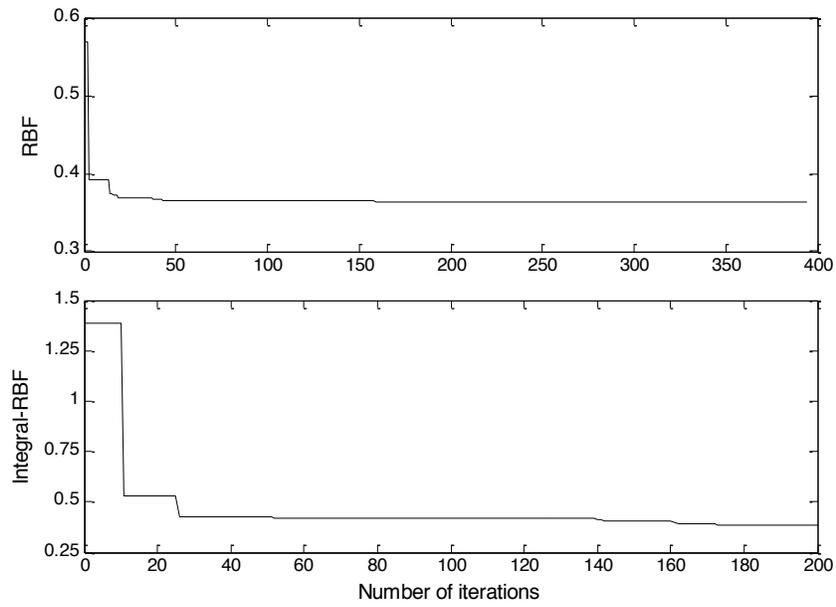

Fig. 3. Convergence of IAE values for RBF and IRBF AQM controllers

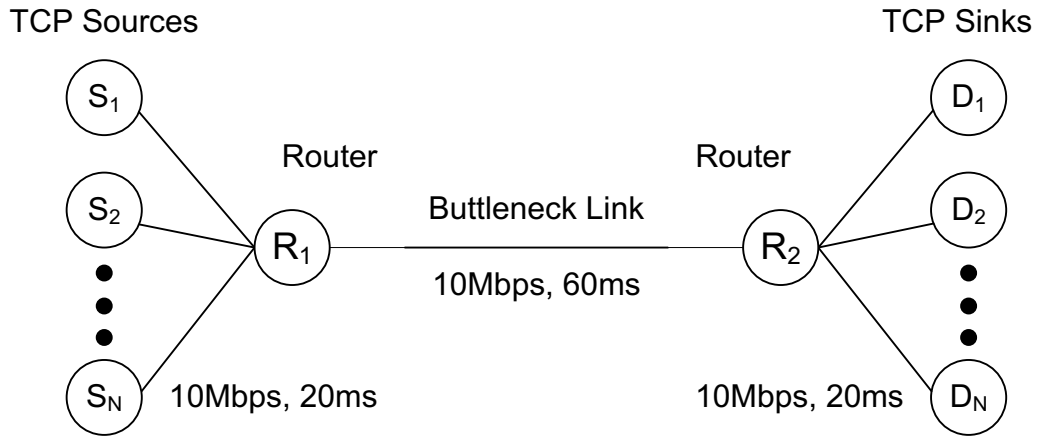

Fig. 4. Dumbbell network topology

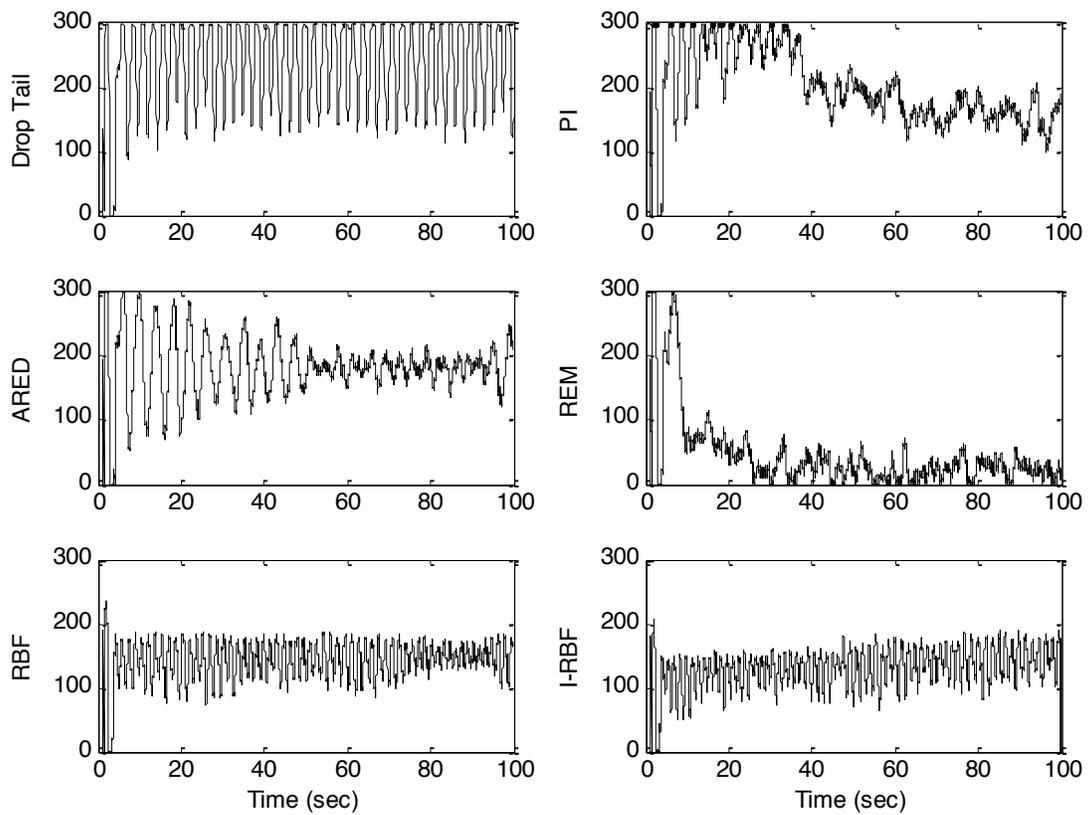

Fig. 5. Queue size (in packets) versus time on dumbbell network for different AQM schemes

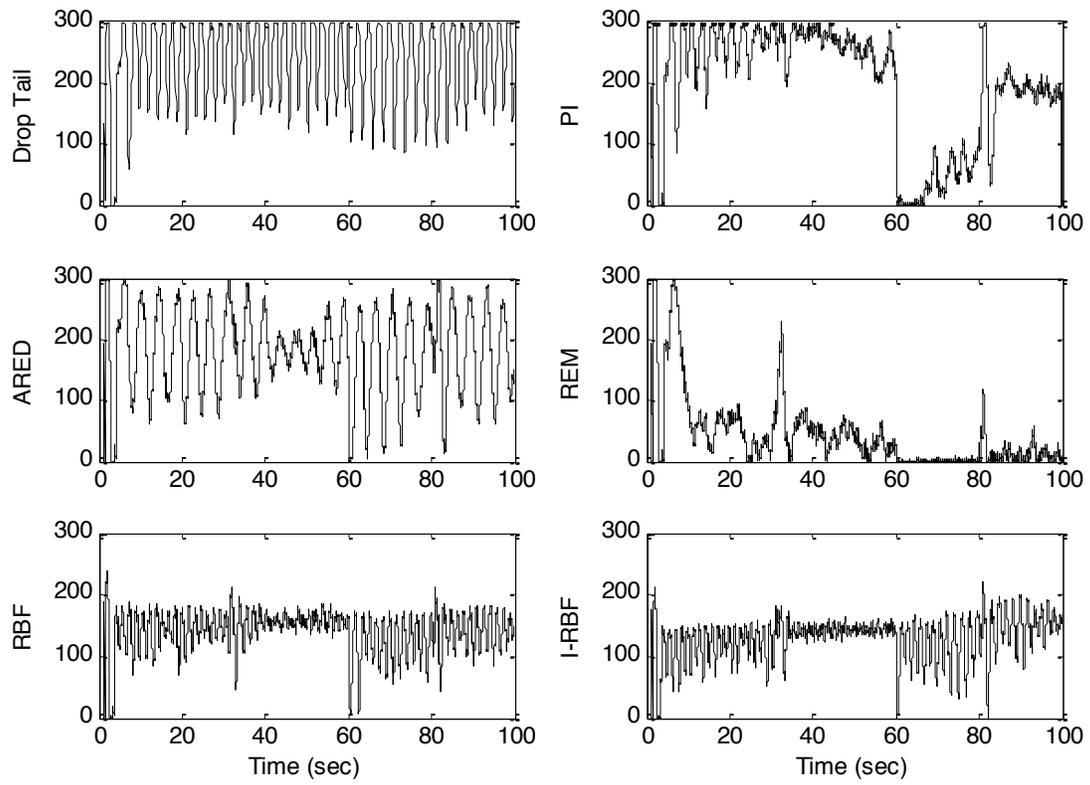

Fig. 6. Queue size (in packets) versus time on dumbbell network subject to dynamic traffic load

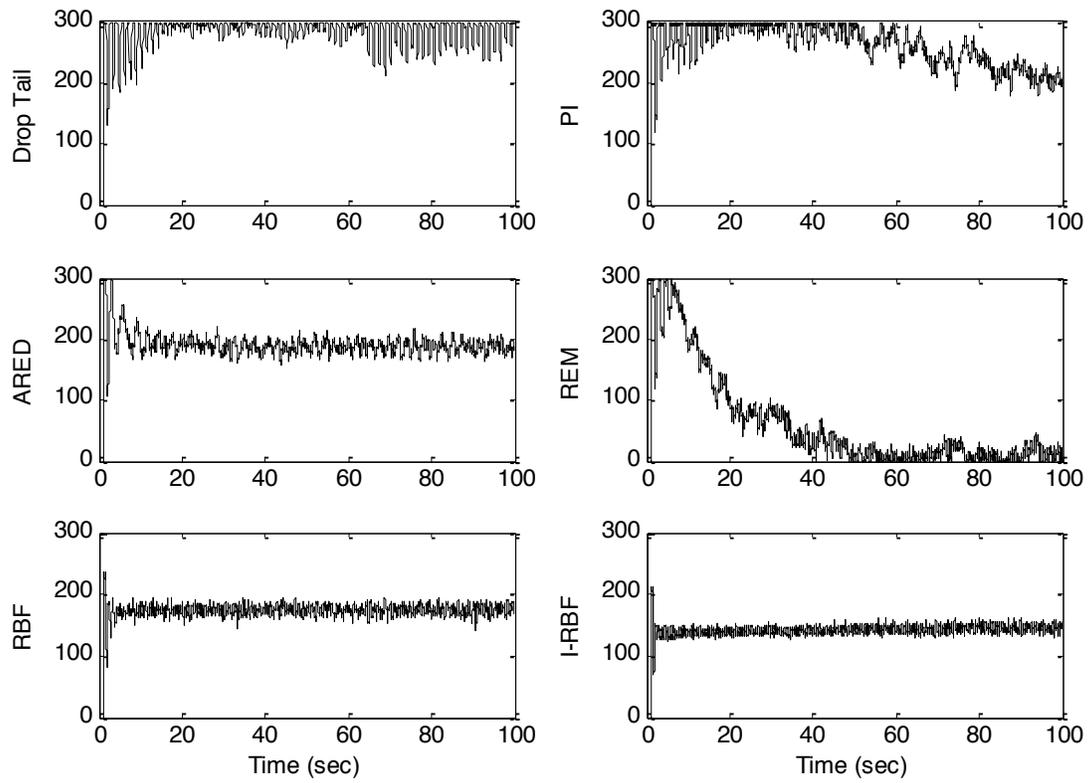

Fig. 7. Queue size (in packets) versus time on dumbbell network with short propagation delay times

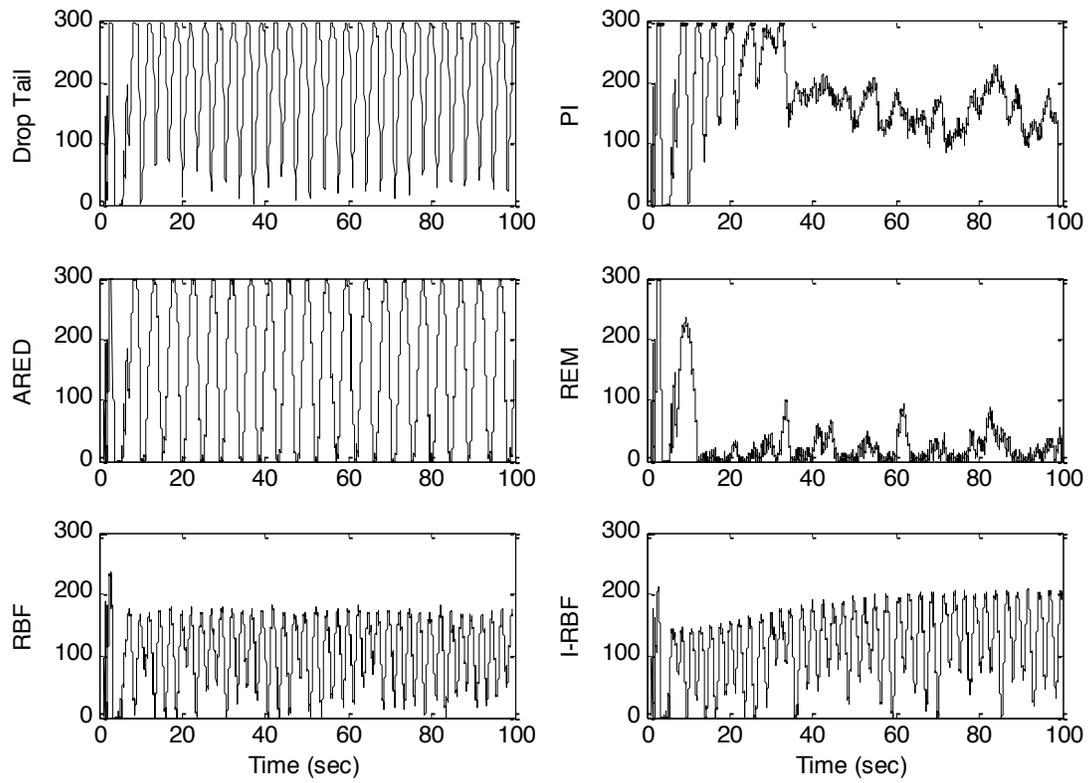

Fig. 8. Queue size (in packets) versus time on dumbbell network with long propagation delay times

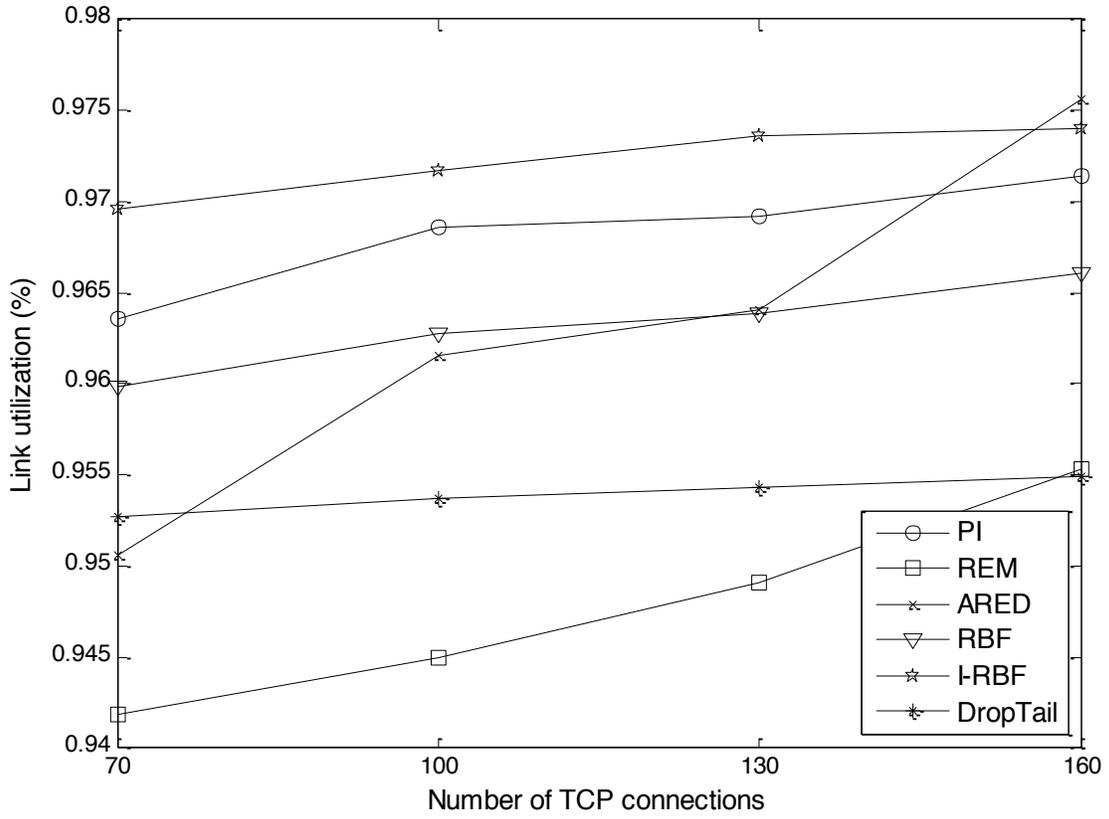

Fig. 9. Link utilization versus number of users for different AQM schemes

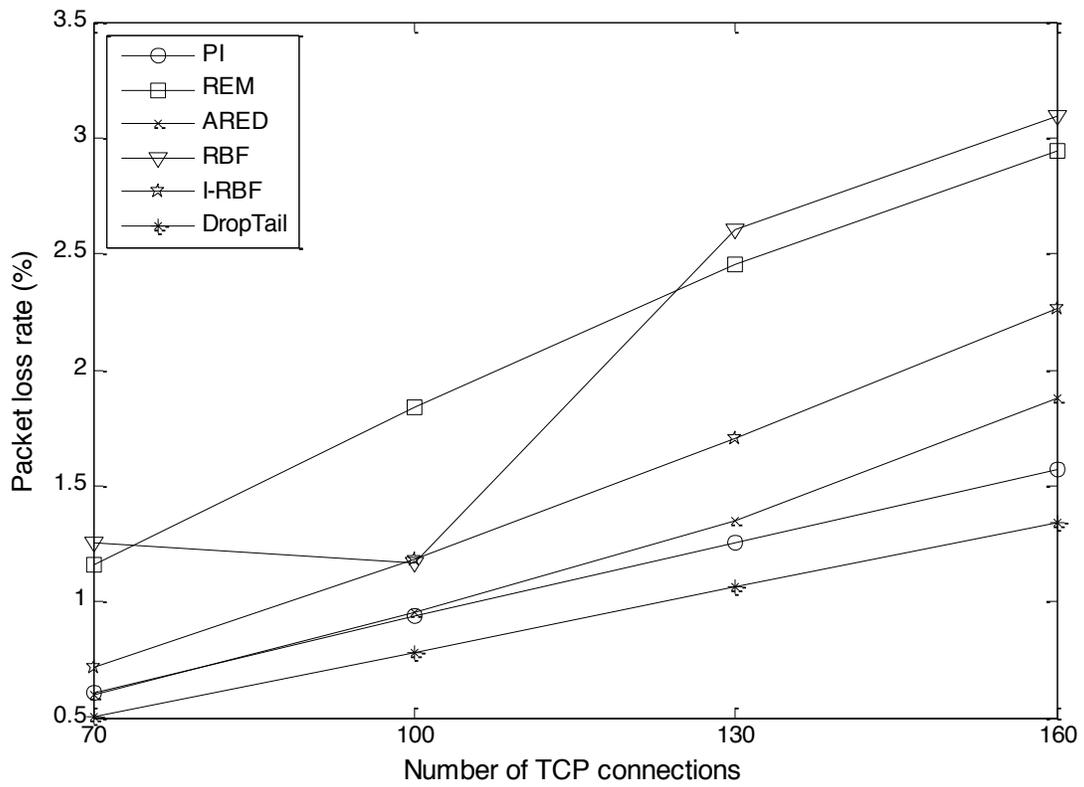

Fig. 10. Packet loss rate versus number of users for different AQM schemes

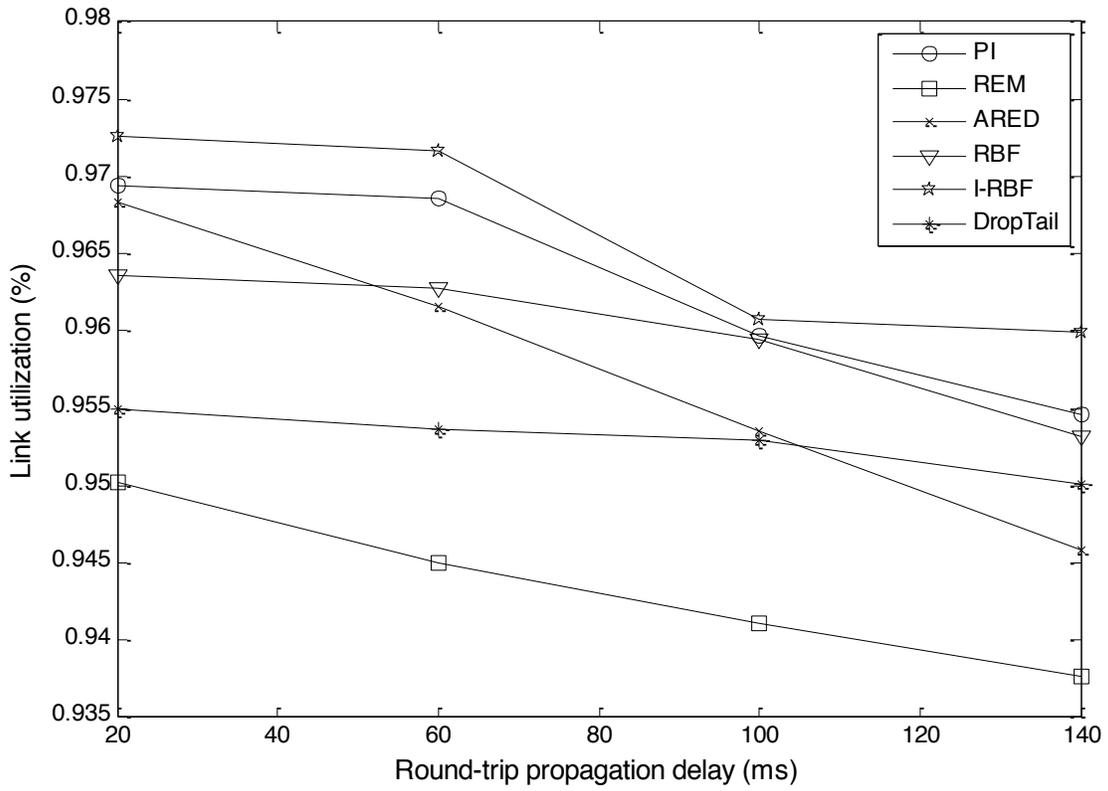

Fig. 11. Link utilization versus propagation delay for different AQM schemes

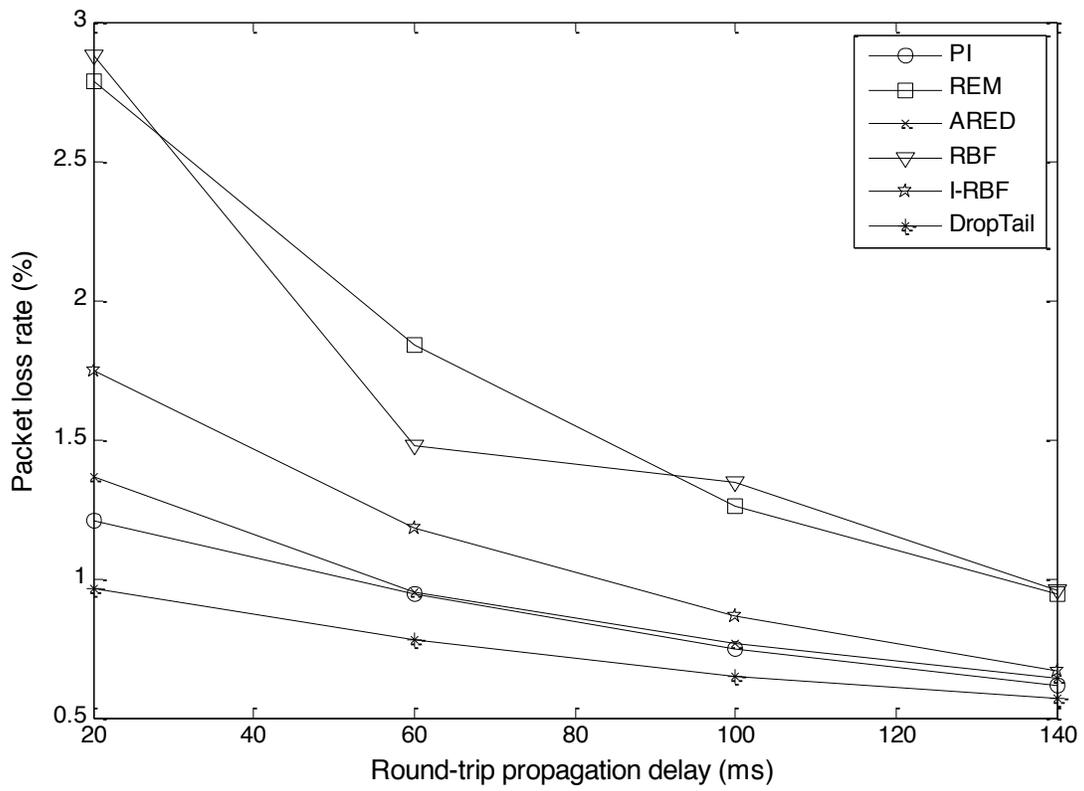

Fig. 12. Packet loss rate versus propagation delay for different AQM schemes